\documentclass[twocolumn]{aastex631}
\usepackage{color}
\usepackage{makecell}
\usepackage{appendix}

\usepackage{amsmath}

\shorttitle{GRB 221009553}
\shortauthors{Xin-Ying Song et al.}

\graphicspath{{./}{figures/}}
\begin{document}

\title{GRB 221009A with an unconventional precursor: a typical two-stage collapsar scenario?}

\correspondingauthor{Xin-Ying Song, Shuang-Nan Zhang}
\email{songxy@ihep.ac.cn, zhangsn@ihep.ac.cn}

%%
%% While authors can be grouped inside the same \author and \affiliation
%% commands it is better to have a single author for each. This allows for
%% one to exploit all the new benefits and should make book-keeping easier.
%%
%% If done correctly the peer review system will be able to
%% automatically put the author and affiliation information from the manuscript
%% and save the corresponding author the trouble of entering it by hand.

\author[0000-0002-2176-8778]{Xin-Ying Song}
\affiliation{Key Laboratory of Particle Astrophysics, Institute of High Energy Physics, Chinese Academy of Sciences, Beijing 100049, China}
\affil{University of Chinese Academy of Sciences, Chinese Academy of Sciences, Beijing 100049, China}
\author{Shuang-Nan Zhang}
\affil{Key Laboratory of Particle Astrophysics, Institute of High Energy Physics, Chinese Academy of Sciences, Beijing 100049, China}
\affil{University of Chinese Academy of Sciences, Chinese Academy of Sciences, Beijing 100049, China}

\begin{abstract}
 As the brightest Gamma-Ray burst (GRB) ever detected, GRB 221009A may offer a chance that reveals some interesting features which are hidden in those bursts that are not so bright. There seems a very weak emission with a flux of $10^{-8}\sim10^{-7}$ erg cm$^{-2}$ s$^{-1}$ between the first pulse ($T_0\sim T_0+50$~s, $T_0$ is the trigger time) and the main burst (appears from $T_0+180$ s). Thus the gap time between them is not really quiescent, and the first pulse could be taken as an unconventional precursor, which may provide a peculiar case study for the GRB-precursor phenomena. A two-stage collapsar scenario is proposed as the most likely origin for this burst. In this model, the jet for the precursor is produced during the initial core-collapse phase, and should be weak enough not to disrupt the star when it breaks out of the envelope, so that the fallback accretion process and the forming of the disk could continue. We present an approach in which the duration and flux both provide constraints on the luminosity ($L_{\rm j}$) and the Lorentz factor at the breakout time ($\Gamma_{\rm b}$) of this weak jet. The estimated $L_{\rm j}\lesssim 10^{49}$ erg s$^{-1}$ and $\Gamma_{\rm b}$ has an order of ten, which are well consistent with the theoretical prediction. Besides, the weak emission in the gap time could be interpreted as a MHD outflow due to a magnetically driven wind during the period from the proto-neutron star phase to forming the accretion disk in this scenario.

 %We propose a possible origin for the first pulse and present an approach in which the duration and flux both provide constraints on the jet luminosity in the scenario of an initial collapsar. For the first pulse, the corresponding jet has a luminosity $\lesssim 10^{49}$ erg s$^{-1}$ and a mild Lorentz factor of several tens, which are well consistent with that it is launched by a weak jet from the initially collapsed core. Besides, the weak emission in the gap time could be interpreted as a MHD outflow due to a magnetically driven wind during the period from the proto-neutron star phase to forming accretion disk. Therefore, a two-stage collapsar scenario is proposed as the most likely origin for the whole burst.
\end{abstract}
%TC:endignore

\keywords{ gamma-ray bursts:individual--radiation mechanisms: non-thermal}


\section{Introduction} \label{sec:intro}

\begin{figure*}
\begin{center}
 \centering
   \includegraphics[width=0.75\textwidth]{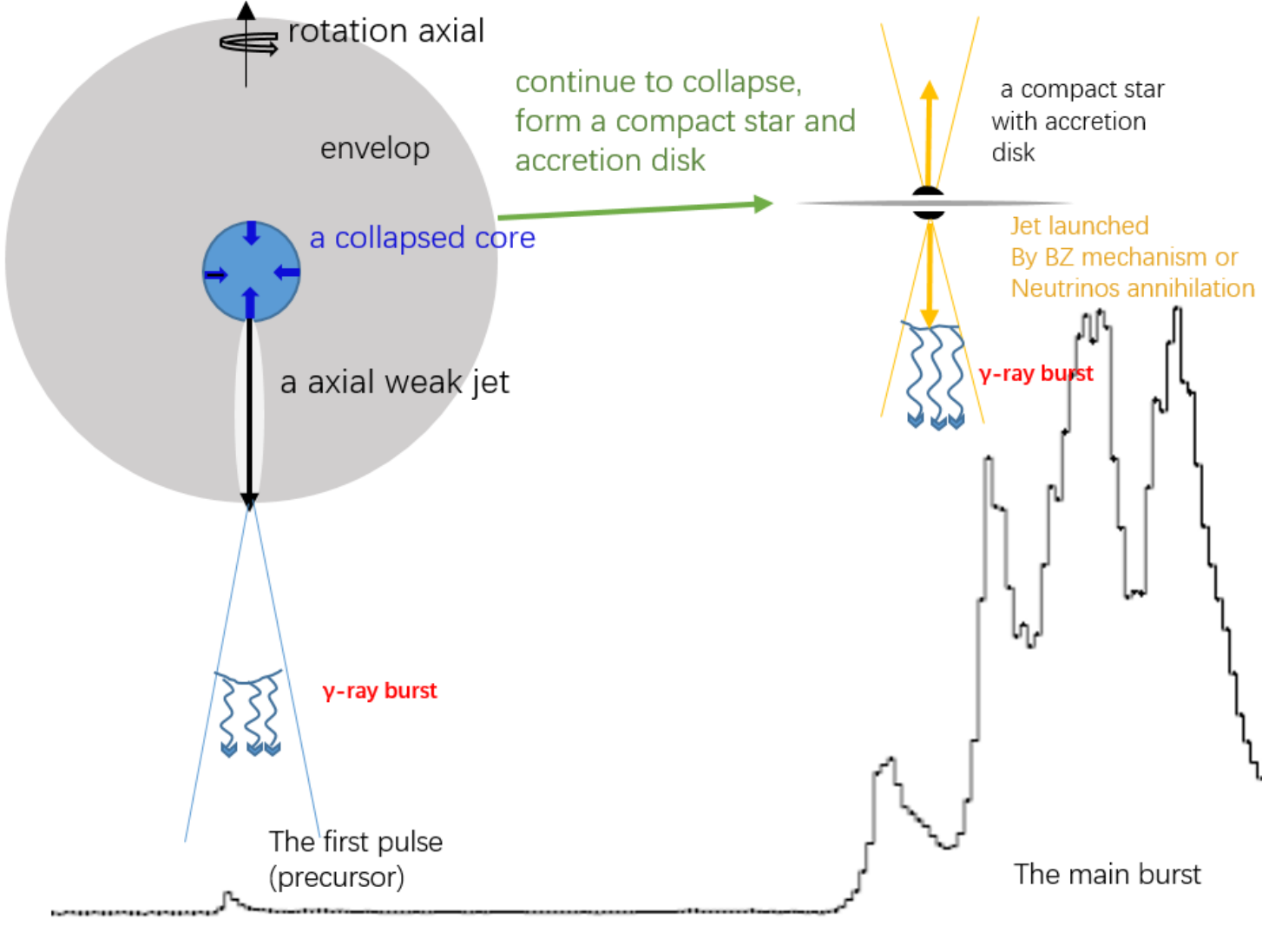}\put(-210,210){`two-stage' model}\\
\caption{A scenario for the `two-stage' model of precursors. 
\label{fig:centralengine} }
\end{center}
\end{figure*}

 Precursors are usual for bright, long GRBs ~\citep[$\sim20\%$, e.g.,][]{2005MNRAS.357..722L} and \textbf{the emission types and jet compositions of their precursors and main bursts are listed in Table~\ref{tab:jet_p_mB}. Quasi-thermal (QT) component could be observed in precursors, as shown in Types 2, 3 and 4, while most of the precursors are found to be non-thermal (NT)\citep[e.g.][]{2022ApJ...928..152L}.} There are some models or mechanisms for precursors of long bursts. Fireball-internal shock (IS) models~\citep[e.g.,][]{2000ApJ...530..292M,2002MNRAS.331..197R,2007ApJ...670.1247W} and jet-cocoon interaction~\citep[e.g.,][]{2017ApJ...834...28N} both predict a precursor with a quasi-thermal (QT) component, as shown in Types 2, 3 and 4 in Table~\ref{tab:jet_p_mB}; the quiescent time for the former is estimated to be about 10 s. Jet-cocoon interaction mechanism and the `two-stage' model~\citep[e.g.,][]{2001APh....16...67C,2007ApJ...670.1247W} both correspond to a scenario of a collapsar. In the `two-stage' scenario, the precursor is from a weak jet which may be produced by a collapsed core~\citep[ e.g., LeBlanc \& Wilson jet,][]{1970ApJ...161..541L} or by a rotating proto-neutron star (PNS) during the initial core-collapse phase, while the quiescent time $\sim$100 s is the timescale of fallback and forming a proto-compact star with an accretion disk; the central engine of the main burst is a black hole (BH) or neutron star (NS). The process of the `two-stage' model is shown in Figure~\ref{fig:centralengine}. In the `magnetar-switch' model~\citep[][]{2013ApJ...775...67B}, the precursor and the main burst arise from accretion of matter onto the surface of the magnetar; the accretion process can be halted by the centrifugal drag exerted by the rotating magnetosphere onto the infalling matter, allowing for multiple precursors and very long quiescent times. Lipunov's works~\citep[e.g.][]{2007ApJ...665L..97L,2009MNRAS.397.1695L} suggest a collapsing `spinar' similar to the `two-stage' model, without any accretion in the process. The origins for precursors are still under debate in some works~\citep[e.g.][]{2005MNRAS.357..722L,2009A&A...505..569B,2018ApJ...866...97B,2022ApJ...928..152L}, and it is of importance to perform the precursor research to understand the physical mechanisms of the GRB central engine.

In this analysis, the so-called precursor in GRB 221009A is not conventional because the `quiescent' time is not really quiescent. Note that in the former works~\cite[e.g.][]{2009A&A...505..569B}, a time interval during which the background
subtracted light curve is consistent with zero, is defined as a `quiescent' time. For GRB 221009A, there exist some weak emissions between the first pulse ($T_0\sim T_0+50$~s, $T_0$ is the trigger time) and the main burst beginning at $T_0+180$~s, as shown in Figure~\ref{fig:middleLC} (a) and (b). However, we still could use the mechanisms for precursors to interpret its origin.

GRB 221009A was detected by many missions, such as Fermi/GBM~\citep[GCN 31565,][]{2022GCN.31565....1L}, Fermi-LAT~\citep[GCN 32637,][]{2022GCN.32637....1B}, Swift/BAT/XRT~\citep[GCN 32632,][]{2022GCN.32632....1D}, Konus-Wind~\citep[GCN 31604,][]{2022GCN.31604....1S}, Insight-HXMT~\citep[ATel 155660,][]{2022ATel15660....1T}, HEBS~\citep[GCN 32751,][]{2022GCN.32751....1L} and LHAASO~\citep[GCN 32677,][]{2022GCN.32677....1H}. For GRB 221009A, the prompt emission has a long duration $\sim 1000$ s. The isotropic-equivalent radiated energy $E_{\rm iso,\gamma}\sim 10 ^{55}$ erg has been reported in some works \citep[e.g.,][]{2023arXiv230213383F,an2023insighthxmt}. Note that the jet of the main burst is highly collimated with a small opening angle $\theta\sim 1.0 ^{\circ}$ \citep{2022arXiv221010673R, an2023insighthxmt}, thus, the outflow has a total energy of $f_{\rm b} E_{\rm iso}\sim 10^{51\sim52}$ erg with $f_{\rm b}\sim\theta^2/2$ and $E_{\rm iso}=E_{\rm iso,\gamma} + E_{\rm iso,\rm k}$, where $E_{\rm iso,\rm k}$ is the isotropic-equivalent kinematic energy and has an order of $10^{55}$ erg. This is a typical released energy for a collapsar to form a BH or a magnetar~\citep[e.g.,][]{2001ApJ...550..410M}, though GRB 221009A is the brightest ever detected in terms of $E_{\rm iso,\gamma}$.

The paper is organized as follows. In Section~\ref{sec:precursor}, we extract the observational properties of the first pulse ($T_0\sim T_0+50$ s) and the followed weak emissions ($T_0+50$ s $\sim$ $T_0+170$ s); in Section~\ref{sec:originfor1stpulse} several scenarios for the precursor and jet launching are discussed. In Section~\ref{sec:discussion}, a conclusion and summary are given based on the discussion.

\begin{deluxetable*}{l|c|c|c|c|c}
\tabletypesize{\footnotesize}
%\tablewidth{0pt}
\tablecaption{The  emission types of long GRBs with precursors. QT: quasi-thermal; NT: non-thermal.\label{tab:jet_p_mB}}
\tablehead{ 
\colhead{Type No.}
&\colhead{1} &\colhead{2} &\colhead{3} &\colhead{4}&\colhead{5} 
}
\startdata
precursor &NT &QT &QT+NT &QT &NT \\
main burst(or following episodes) &NT &NT &NT &QT+NT &QT+NT \\
%observed  &$\surd$ &$\surd$ &$\surd$ &$\surd$ &? \\
\enddata
\end{deluxetable*}

\section{The observational properties of the first pulse and weak emissions}\label{sec:precursor}
Background (BG) estimation for extremely long GRB 221009A is important. We use the data from nearby orbit as BG for GBM NaI 8 detector, and polynomial with 0-2 orders for GBM BGO 0 detector above 385 keV. The details are shown in Appendix~\ref{sec:bg}. As shown in Figure~\ref{fig:middleLC} (a) and (b), some weak emissions exist between the first pulse and the main burst, and are mainly from the lower energy band ($\lesssim 100$ keV).
The first pulse lasts $\sim 50$ s. After a very weak emission with duration of around 70 s, a long bump of 60 s comes before the main burst as shown in Figure~\ref{fig:middleLC} (c).  

 \begin{figure*}
\begin{center}
 \centering  
 \includegraphics[width=.88\textwidth]{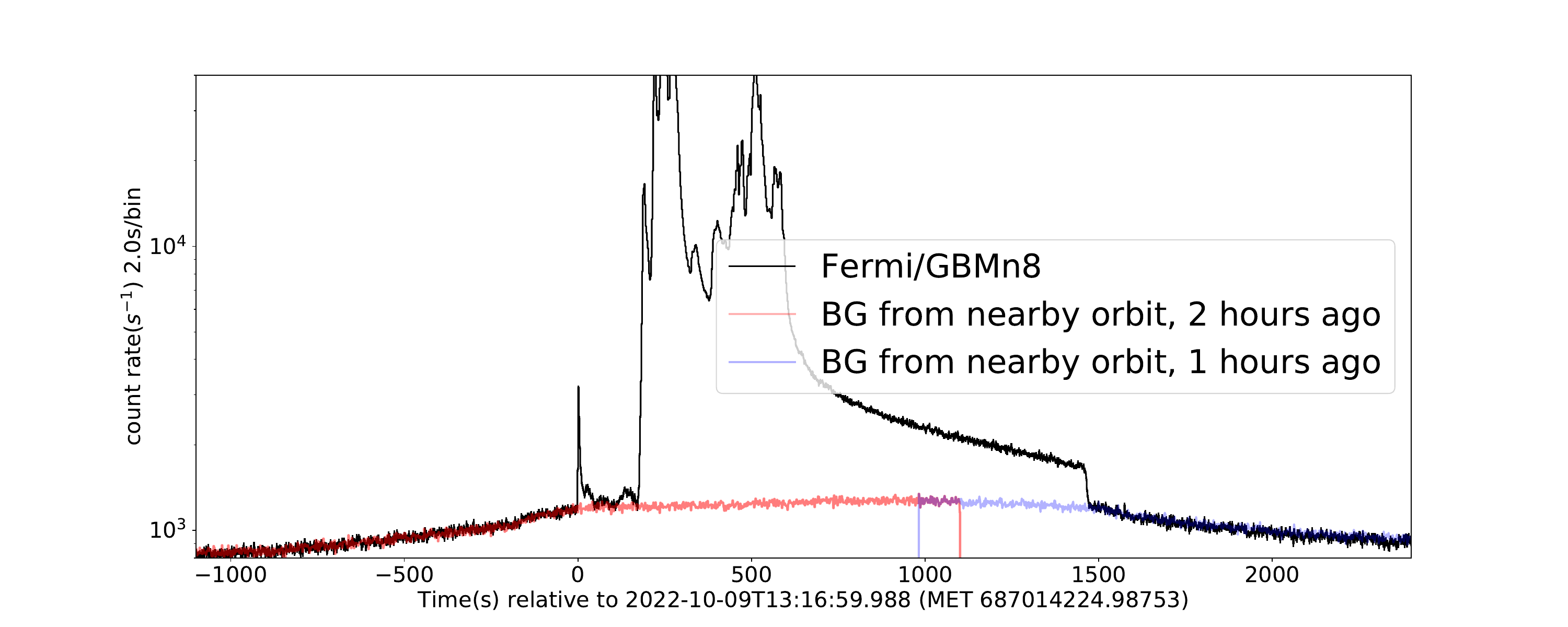}\put(-130,140){(a) NaI 8}\\
   \includegraphics[width=.88\textwidth]{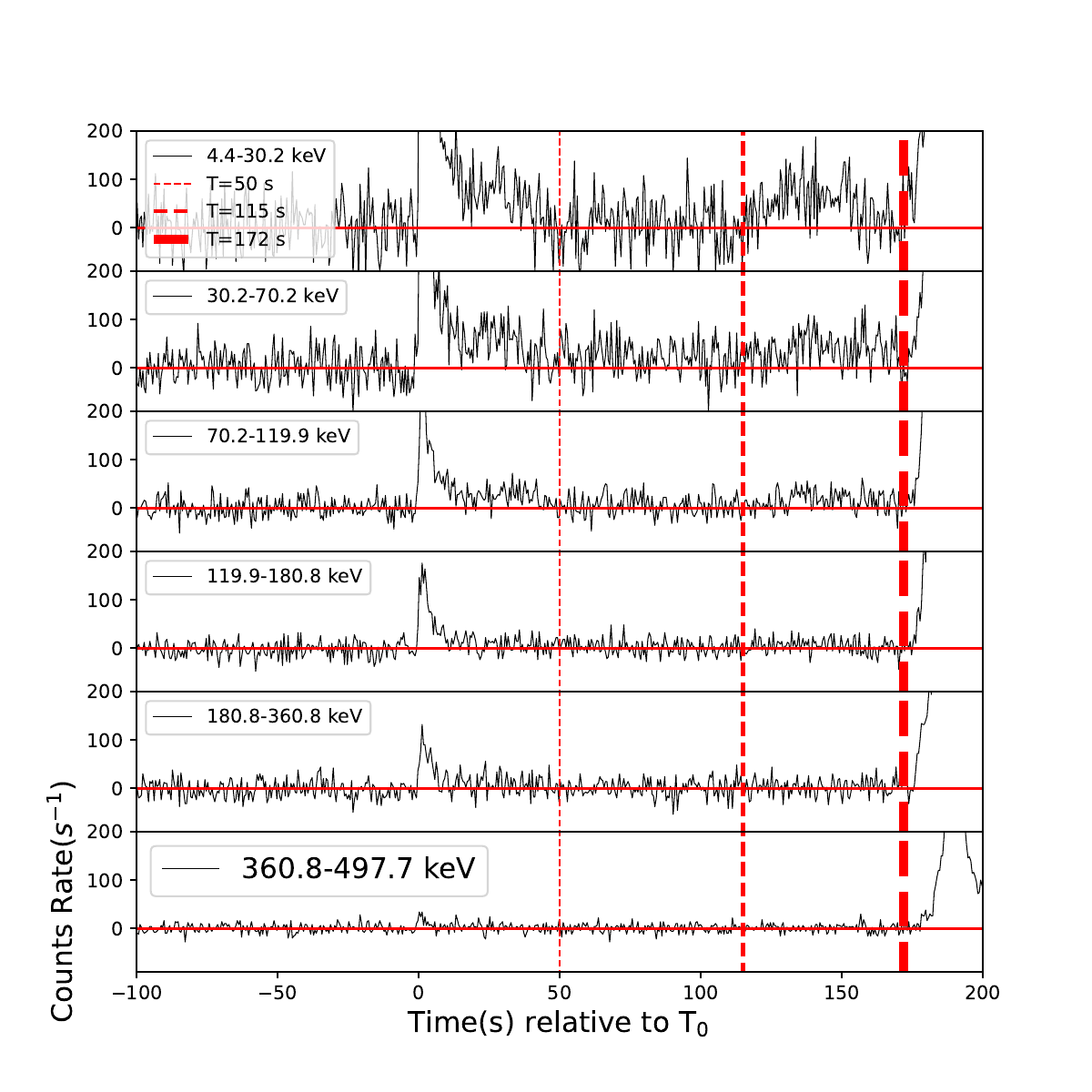}\put(-200,380){(b) NaI 8}\\
 \caption{ (a) The light curve and BG shape from the data of NaI 8.  (b) The BG subtracted light curves in different energy bands of NaI 8. (c) The light curve from $T_0$ to $T_0+200$ s. (d) The spectrum of the first pulse. (e) The light curves from GBM NaI 8 detector and HXMT, $\alpha$ and $E_{\rm p}$ values of precursor. (f) The spectrum of $T_0+50$ s to $T_0+115$ s. (g) The spectrum of $T_0+115$ s to $T_0+172$ s.
\label{fig:middleLC} }
\end{center}
\end{figure*} 

\begin{figure*}
\begin{center}
 \centering 
   \includegraphics[width=.88\textwidth]{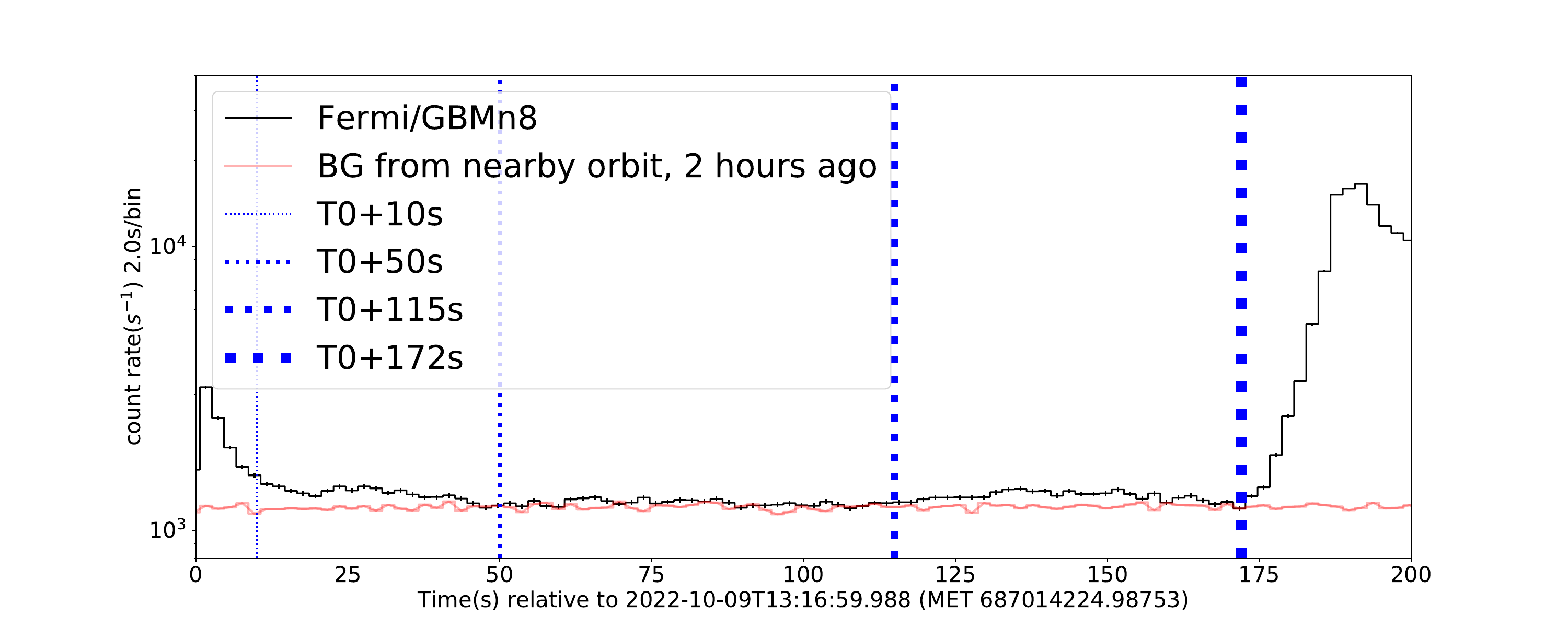}\put(-130,140){(c)}\\
   \includegraphics[width=0.5\textwidth]{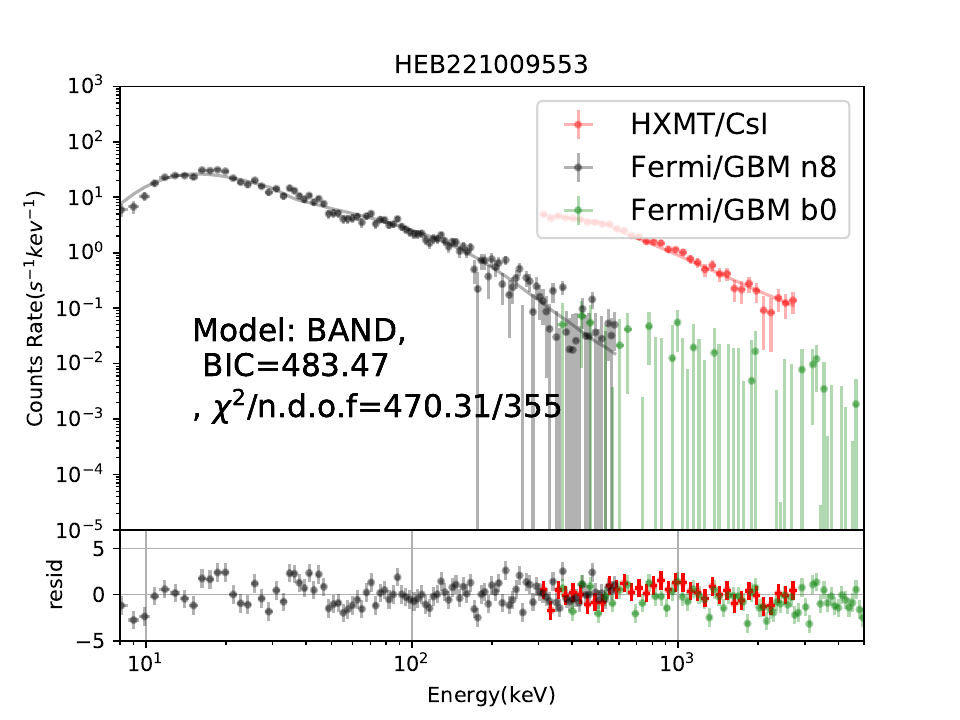}\put(-190,150){(d) $T_0$ to $T_0+10$ s }
   \includegraphics[width=0.5\textwidth]{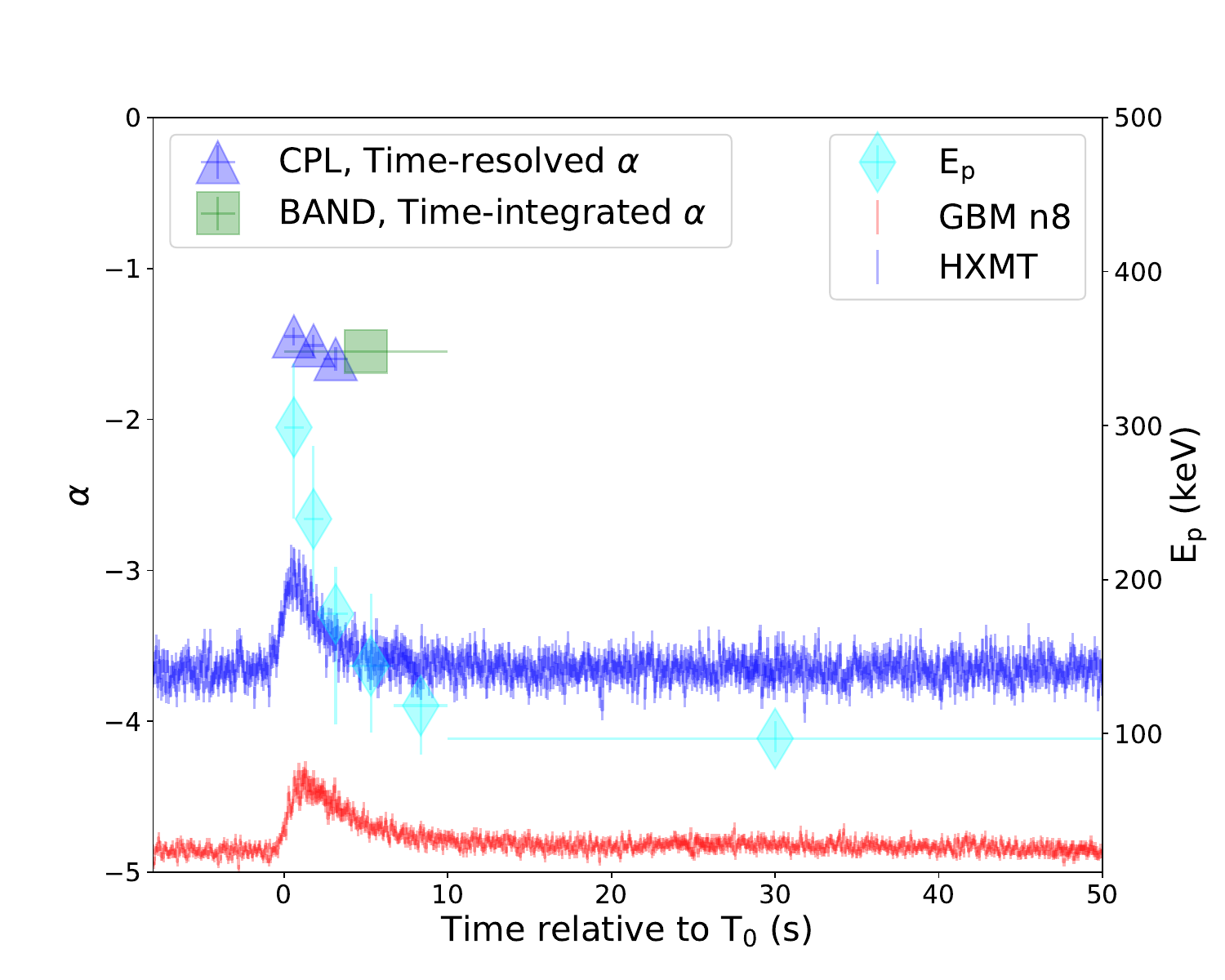}\put(-140,140){(e) }\\
   
   \includegraphics[width=0.5\textwidth]{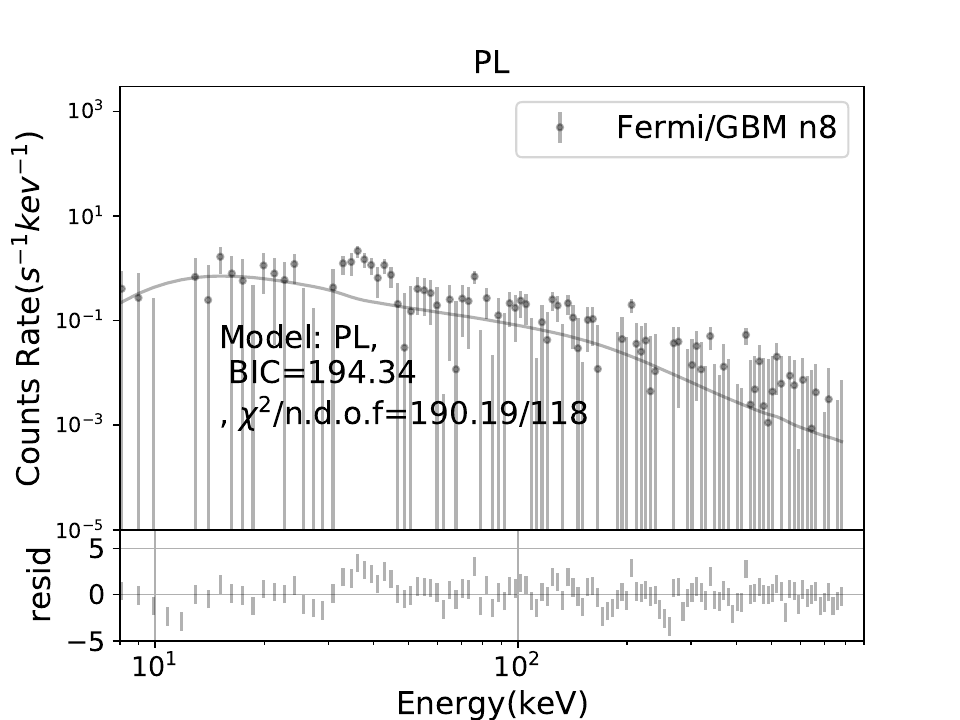}\put(-180,140){(f) $T_0+50$ s to $T_0+115$ s}
   \includegraphics[width=0.5\textwidth]{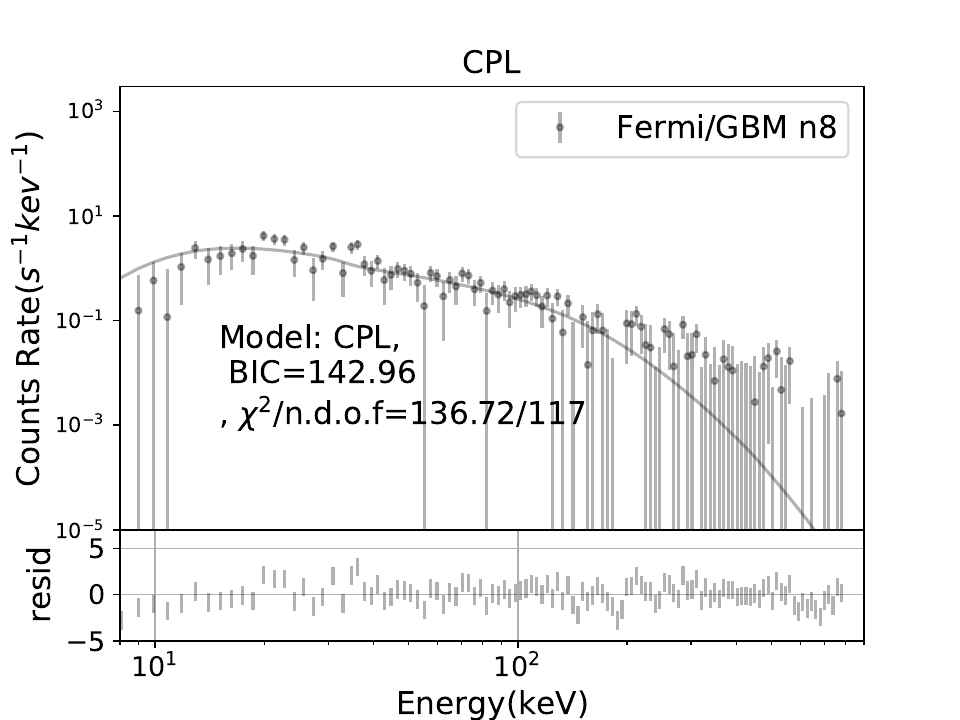}\put(-140,140){(g) $T_0+115$ s to $T_0+172$ s}\\
\center{Figure. \ref{fig:middleLC}--- Continued.} 
\end{center}
\end{figure*}

\subsection{The first pulse from $T_0$ to $T_0+50$ s  }\label{subsec:obs_precursor}
 Fittings with BAND, exponential cut-off power law (CPL), and power law (PL) function\footnote{ The formulae for spectral models, BAND, CPL, PL could be found in the Appendix in \cite{2022MNRAS.517.2088S}.} are performed on the time-integrated spectrum from $T_0$ to $T_0$+10 s which contains $80\%$ photons. The Markov Chain Monte Carlo (MCMC) fitting is performed to find the parameters with the maximum Poisson likelihood. The BAND model is determined to be the best model by the method of bayesian information criterion~\citep[BIC,][]{2016MNRAS.463.1144W}, and require $\Delta$BIC is at least 6\footnote{$\Delta$BIC$\geq10$, the preference is very strong; $10>\Delta$BIC$\geq6$, strong; $6>\Delta$BIC$\geq2$, positive. }. If with the contribution from HXMT/CsI detectors which have large effective area in high energy region~\citep{2022ApJS..259...46S}, the low energy photon index ($\alpha$), the high energy photon index ($\beta$) and the peak energy ($E_{\rm p}$) of $\nu F_{\nu}$ spectrum is determined to be $\alpha=-1.55\pm0.03$, $\beta= -2.02\pm0.02$ and $E_{\rm p}=242.9\pm 113.0$ keV as shown in Figure~\ref{fig:middleLC} (d).
 The low photon energy index $\alpha<-2/3$~\citep[the so-called `line of death',][]{Preece_1998}, which is consistent well with the synchrotron mechanism.  

The constant cadence~\citep[CC,][]{2014On} method and Bayesian blocks~\citep[BBlocks,][]{2013ApJ...764..167S} method with a false alarm probability $p_0$= 0.01 are used for binning in time-resolved analysis. We also require the signal-to-noise ratio (S$/$N)~$\geq$30 at least in one detector, so we combine some adjacent bins. The time bins are [0., 1.2], [1.2, 2.4], [2.4, 3.9], [3.9, 6.7], [6.7, 10] s. The evolution of $\alpha$ is shown in Figure~\ref{fig:middleLC} (e). Generally, the double-tracking trend of $\alpha$-flux and $E_{\rm p}$-flux, is observed in the first 10 s, which is consistent well with that of one-zone synchrotron model~\citep[e.g.,][]{2014NatPh..10..351U,2019ApJ...884..109L}. Note that $\alpha<-1$ for the first pulse, implies non-thermal (NT) emission is dominant. The internal-collision-induced magnetic reconnection and turbulence mechanism ~\cite[ICMART,][]{2011ApJ...726...90Z} is preferred as the one-zone synchrotron model for this NT emission mainly.

\subsection{The weak emissions between the the first pulse and the main burst}
The emission from $T_0+50$ s to $T_0+115$ s has S/N$\sim10$ in NaI 8 detector. Therefore, it is difficult to describe the shape of spectrum. The observed flux is estimated to be $\sim10^{-8}$ erg cm$^{-2}$ s$^{-1}$ with PL model as shown in Figure~\ref{fig:middleLC} (f). The long bump from $T_0+115$ s to $T_0+172$ s is best described by CPL function with $\alpha=-1.00\pm0.15$ and $E_{\rm p}=78.5\pm 8.0$ keV. The flux is $\sim 10^{-7}$ erg cm$^{-2}$ s$^{-1}$ as shown in Figure~\ref{fig:middleLC} (g).

\section{the possible origin of the first pulse}\label{sec:originfor1stpulse}

%\cite{2007ApJ...670.1247W} have describe a scenario in which a precursor caused by a weak axial jet followed by a main burst after hundreds seconds. This weak jet has two origins: it could be a axial jet predicted by LeBlanc \& Wilson \citep[LW jet, e.g., ][]{1970ApJ...161..541L}); alternatively, a Poynting flux might be generated as the proto-neutron star (PNS) spin up. For both origins, the jet propagates within the stellar atmosphere and creates a cocoon; then the jet head moves outward and finally breaks out at the stellar envelope. At this time, two component of the cocoon, including the jet material and a stellar material that was shocked by the expanding high pressure cocoon~\citep{2017ApJ...834...28N} contribute to the ejecta.      
%For the common case in which the opening angle of the main burst is not so small or the burst has higher $z$, if the observation is not very head-on, the weak emission even the bump may be missed by the detector and the first pulse is mistaken to be a precursor with a time gap of hundreds of seconds from the main burst. 
There are some common characteristics between the first pulse in GRB 221009A and the conventional precursor: much weaker than the main burst and 
a gap time from the main burst. Thus, several models or mechanisms for precursors could be used to interpret origins of the first pulse as well. Fireball-internal shock (IS) models and jet-cocoon interaction are excluded first, because
there seems not any evident QT component in the emission of the precursor as discussed in Section~\ref{subsec:obs_precursor}.

\textbf{Besides, the gap time between the precursor and the main burst is too long ($\sim 100$ s) for the fireball-IS model. In the fireball-IS model, the gap time is contributed by three parts~\cite[e.g.][]{2007ApJ...670.1247W}.
The first part ($t_1$) is the time that the rarefaction wave takes to arrive at the reverse shock. Once the jet head reaches the stellar surface, the pressure in front of the jet head decreases suddenly, and a rarefaction wave will form and propagate back into the shocked jet material at the speed of sound ($c_{\rm s}=c/\sqrt{3}$). The width of the shocked jet is less than the distance from the core to the envelope ($r$), thus $t_1\lesssim r/c_{\rm s}\simeq6r_{11}$ s. The second part ($t_2$) is from the time that the unshocked jet pass through the envelop, $t_2=r/c=3 r_{11}$ s. The internal shock dissipation occurs at about $R_d$ as the beginning of the main burst, and the third part ($t_3$) is the delay of the main burst to the precursor, $t_3=R_d/2\Gamma^2 c=1.7R_{d,15}\Gamma^{-2}_2$ s. The gap time is the sum of $t_1$, $t_2$ and $t_3$ and has an order of 10 s. }

It seems that the `two-stage', `magnetar-switch' and spinar models could be consistent with the NT emission for the precursor and long gap time. Here we define two quantities: (1) the Lorentz factor $\Gamma_{\rm b}$ at the breakout time of the jet for the precursor, which is the Lorentz Factor when the jet breaks out of the envelope and can be taken as the maximum Lorentz factor of the jet passing through the envelope;\footnote{Note that $\Gamma_{\rm b}$ may be not the maximum Lorentz Factor of the jet ($\Gamma$) and $\Gamma_{\rm b}<\Gamma$;} (2) the luminosity of the jet ($L_{\rm j}$) for the precursor. In the `magnetar-switch' and spinar model, $\Gamma_{\rm b}$ and $L_{\rm j}$ are not constrained especially, while for the two-stage model, the jet for the precursor should be weak enough not to disrupt the star, so that the fallback accretion process and the forming of the disk could continue. In details, a mild $\Gamma_{\rm b}<100$ and the released energy $\lesssim10^{50}$ erg are both required~\citep{2007ApJ...670.1247W}.
Therefore, the estimation of $L_{\rm j}$ and $\Gamma_{\rm b}$ for the first pulse is important. The `two-stage' model can be excluded if a weak jet is not consistent with the data. The first pulse of GRB 221009A has a long duration of tens of seconds ($80\%$ photons are in $\sim$10 s). Thus, $t_{b}\lesssim 10$ s, where $t_{b}$ is the time taken by the jet head to move from the interior of the star to the surface.

Assuming the jet acceleration is saturated, we have Equation (12) in \cite{2007ApJ...670.1247W} to describe the relation among $\Gamma_{\rm b}$, $L_{\rm j}$ and $t_{\rm b}$,
\begin{equation}\label{eq:eq1}
   \Gamma_{\rm b}\gtrsim  10r_{11}^{1/2} L_{\rm j,49}^{-1/4} t_{\rm b,10}^{-3/4}, 
\end{equation}
 where $r\sim 10^{11}$ cm is the distance from core to envelope; CGS\footnote{the convention $Q = 10^{n}Q_{n}$ is adopted for CGS units.} units are used here. The flux in the first 10 s is $\sim 2\times10^{-6}$ erg cm$^{-2}$ s$^{-1}$ ($L_{\rm iso, \gamma}\sim 10^{50}$ erg  s$^{-1}$ with redshift $z=0.151$ from \cite{2022GCN.32648....1D}; here $L_{\rm iso, \gamma}$ denotes $E_{\rm iso, \gamma}/T$, $T$ is the duration time in the rest frame of central engine; $T=T_{\rm obs}/(1+z)$ with $T_{\rm obs}$ is that in the rest frame of the observer), thus, the isotropic-equivalent luminosity $L_{\rm iso}$ could be $\gtrsim 10^{50}\sim10^{51}$ erg  s$^{-1}$~\citep[$L_{\rm iso}=L_{\rm iso, \gamma}/\epsilon_{\gamma}$ with radiative efficiency $\epsilon_{\gamma}\sim50\%-90\%$ for ICMART mechanism, ][]{2011ApJ...726...90Z}. Considering the opening angle ($\theta_{\rm b}$) at the breakout time $\sim 1/\Gamma_{\rm b}$ and $L_{\rm j}\sim L_{\rm iso} \theta_{\rm b}^{2}/2$, we have 
 \begin{equation}\label{eq:eq2}
 \Gamma_b\sim(2L_{\rm j}/ L_{\rm iso})^{-1/2}.
 \end{equation}
 
\begin{figure*}
\begin{center}
 \centering
   \includegraphics[width=0.65\textwidth]{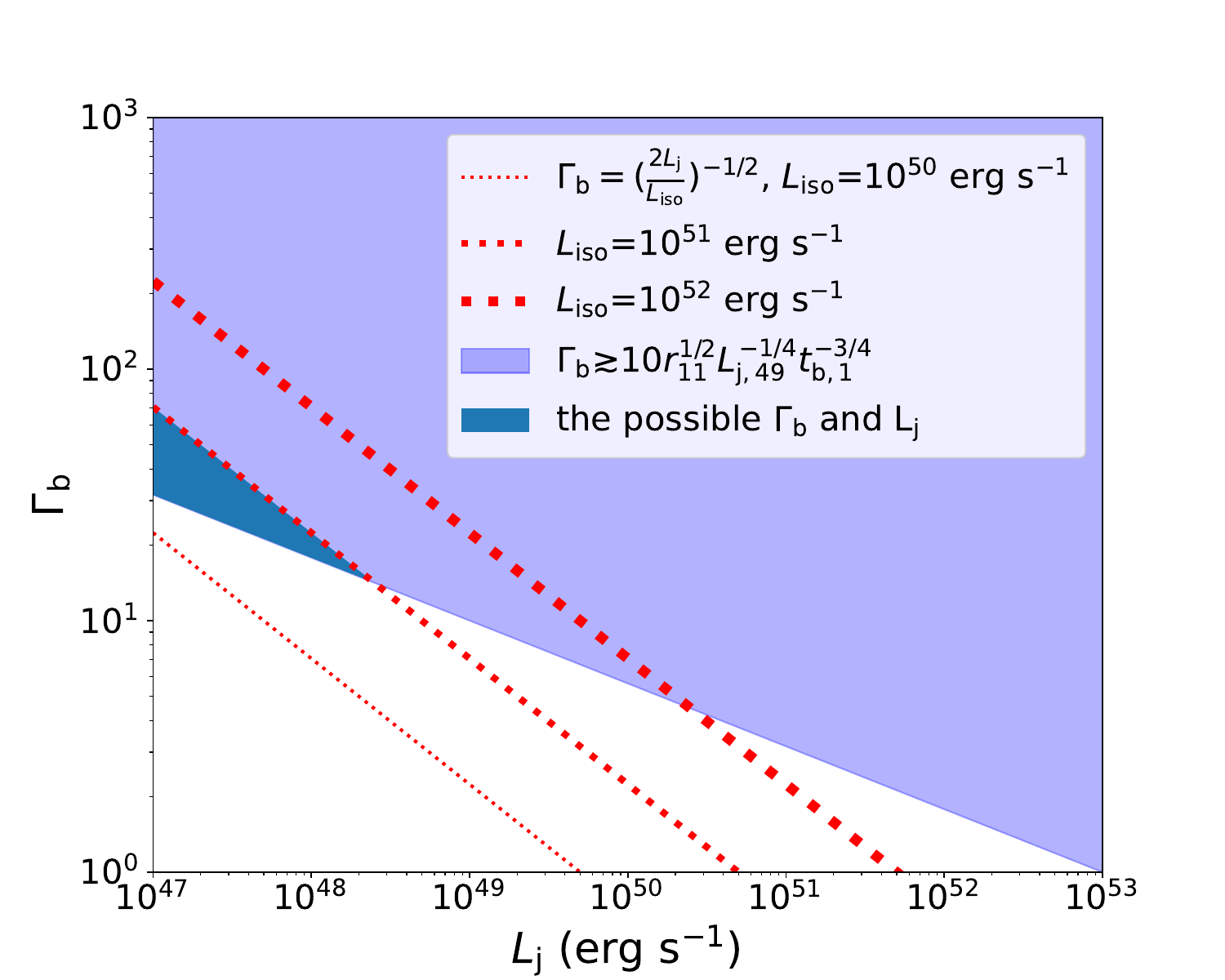}\put(-120,80){$t_{\rm b}\lesssim 10s$}
\caption{The relation between $\Gamma_{\rm b}$ and $L_{\rm j}$ in Equation~(\ref{eq:eq2}) in red dotted lines of different $L_{\rm iso}$. The violet shadow denotes the possible $\Gamma_{\rm b}-L_{\rm j}$ only with the Constraint~(\ref{eq:eq1}) if $t_{\rm b}\lesssim 10$ s. The dark blue shadow denotes the region of possible $\Gamma_{\rm b}$ and $L_{\rm j}$.\label{fig:Gamma_b_Lj}}
\end{center}
\end{figure*}

Here, we present an approach to limit $L_{\rm j}$ and $\Gamma_{\rm b}$. By combining the above equation and constraint, the region for the possible values for $L_{\rm j}$ and $\Gamma_{\rm b}$ is the overlap of these two as shown in Figure~\ref{fig:Gamma_b_Lj} in dark blue. The range of $L_{\rm j}$ corresponds to $\lesssim10^{48}$ erg s$^{-1}$, and $\Gamma_b$ has an order of 10, which is consistent well with the weak jet assumption. Note that this estimation is approximate with the orders of magnitudes of these quantities, e.g., $r_{11}\sim 1$ and $L_{\rm iso,51}\sim 1$. For specific values, the order of magnitudes of the results will not be changed much. If we use the full time of the first pulse, $t_{b}\lesssim 50$ s, the edge of the violet shadow in Figure~\ref{fig:Gamma_b_Lj} which denotes the lower limit of the $\Gamma_{\rm b}$ will move to the lower range according to Constraint~(\ref{eq:eq1}); $L_{\rm iso,\gamma}\sim 2.5\times 10^{49}$ erg s$^{-1}$ is smaller, thus the possible $\Gamma_{\rm b}$ and $L_{\rm j}$ become much smaller that those with $t_{b}\lesssim 10$ s.

For the unsaturated acceleration case, as already calculated in Equation (11) in \cite{2007ApJ...670.1247W} for $t_{\rm b}\lesssim10$ s, $\Gamma_{\rm b}\gtrsim10$. In this case, the upper limit of $\theta_{\rm b}\lesssim0.1$, so that $L_{\rm j}\sim L_{\rm iso} \theta_{\rm b}^{2}/2$ is still weak with luminosity of $ 10^{48}\sim10^{49}$ erg s$^{-1}$, since $L_{\rm iso}\sim 10^{50}-10^{51}$ erg s$^{-1}$.

Therefore, from the above discussion, the observed flux and duration of this pulse both provide constraints on $L_{\rm j}\lesssim10^{49}$ erg s$^{-1}$ in this case. Otherwise, a shorter duration or larger luminosity could result in a larger $\Gamma_{\rm b}$ or $L_{\rm j}$, so that the assumption of a weak jet fails. Note that the estimated $\theta_{\rm b}$ is not small and the weak jet from the initial collapsar is an axial jet~\cite[e.g.,][]{1970ApJ...161..541L}; the strong jet of the main burst launched by, e.g. the Blandford–Znajek (BZ) mechanism~\citep{1977MNRAS.179..433B}, is also in the axial direction of rotation, thus, we still could see the burst with the first pulse, though the latter is highly collimated with a much smaller opening angle. %Therefore, in this model observationally the main burst is always preceded with a precursor (or the first pulse), but it should happen more frequently that the precursor is observed but the main burst is not.

 \textbf{One question arises that there may be a possible jet-cocoon interaction during the jet passing through the envelop. If $\Gamma_{\rm b}$ is large enough, the mixing between the two components (the jet material and a stellar material shocked by the expanding high pressure
cocoon) could be ignored. As simulated numerically in \cite{2017ApJ...834...28N}, a relativistic cocoon may produce a short (a few seconds)
extremely bright QT burst (with the observed temperature of $10\mbox{--}$100 keV) as the precursor. Otherwise, if $\Gamma_{\rm b}$ is small enough, the partial mixing occurs~\citep{2017ApJ...834...28N}. As a result, optical/UV emission should appear with temperature of $\sim10^4$ K, beyond the observed precursor phase; unfortunately, the optical/UV observations started much later and can not offer more information on this. There is no evident thermal component observed in the precursor, furthermore, most of emission in the precursor is NT at least. This might be also evidence that the jet responsible for the precursor is weak with a mild Lorentz Factor.}

 Another constraint for the origin is the weak emission between the first pulse and the huge main burst. As estimated in Section~\ref{subsec:obs_precursor}, the flux of the weak emission has an order of $10^{-8}\sim10^{-7}$ erg cm$^{-2}$ s$^{-1}$, which is one to two magnitudes smaller than that of the first pulse ($\sim 10^{-6}$ erg cm$^{-2}$ s$^{-1}$). In a collapsar scenario of the `two-stage' model, the `quiescent' time is the time scale of the period from the PNS phase to forming an accretion disk. During this time, the newborn NS would launch a strong neutrino-driven wind, or a magnetically driven wind due to the differential rotation of the NS. A semi-analytical
spindown formula~\citep{2014ApJ...785L...6S} for a magnetically driven wind gives a luminosity of 
\begin{equation}
    L\simeq 10^{48}  B^2_{14} R^3_{6} P^{-1}_{-4}{\rm erg\, s}^{-1},
\end{equation}
where $B$ is the surface magnetic field strength at
the polar cap region, $R$ is the radius of the NS, and $P$ is the period. It seems that this spindown mechanism could produce a MHD outflow with luminosity of one to two magnitudes smaller than that of the first pulse ($\sim10^{49}$ erg s$^{-1}$) if the values of $B$, $R$ and $P$ are in reasonable ranges.

For the `magnetar-switch' model, the longer the waiting time, the higher the stored energy available for the next emission episode. The `quiescent' time for GRB 221009A is long, and the main burst is extremely bright, which seems consistent with the prediction of `magnetar-switch'. There are three mechanisms of energy extraction for a magnetar as a central engine for a GRB, including 1) spin down controlled by magnetic dipole radiation, 2) extracting differential rotational energy of the NS through an erupting magnetic bubbles by winding up the poloidal
magnetic field into the a toroidal configuration~\citep{1998ApJ...505L.113K}, and 3) accretion. \textbf{In the propeller mechanism, the accretion process can be halted by the centrifugal drag exerted by the rotating magnetosphere onto the infalling matter, and during the halting time, there should be no evident emission emitted. The weak emission of 60 s before the main burst is not predicted in this model. If `magnetar-switch' works, it is necessary to interpret the long bump of $60$ s before the main burst at least. If we assume it is not from the beginning of the re-accretion, it must be from the magnetic dipole radiation or the erupting magnetic bubbles. The former should exist during the burst and does not begin at  $T_0+112$ s; the latter occurs at a hot PNS phase, and the released energy ($\sim 10^{51}$ erg) seems too high for the bump. If it is the beginning of the re-accretion, it is not reasonable that it lasts $\sim 60$ s with a very low luminosity ($\sim 10^{49}$ erg s$^{-1}$, corresponding to $\sim 10^{-7}$ erg cm$^{-2}$ s$^{-1}$ at the distance of this GRB) as the next
%flux ($\sim 10^{-7}$ erg cm$^{-2}$ s$^{-1}$) as the next
%应该用辐射的光度，比如$\sim 10^{49}$ erg s$^{-1}$ (具体是多少，需要你根据re-accretion的模型给出），不能用观测到的流量。然后根据光度+距离，得到流量，再和观测的比较，说明观测的结果和模型不符。
emission episode with high energy. Moreover, the magnetar-switch scenario offers a good explanation for these GRBs whose precursors have spectral and temporal properties similar to the main
prompt emission, and smaller, but comparable, energetics ~\citep{2013ApJ...775...67B}, because the origins for precursors and main bursts are the same in this model. It is significant that the energies released in precursor and the main emission are not comparable for GRB 221009A. Therefore, considering these inconsistencies, the `magnetar-switch' model may be not the best interpretation for GRB 221009A.}

In the scenario of the spinar model, the details for a weak jet corresponding to the precursor are not predicted or constrained. It also occurs in a collapsar scenario, thus we think the production of the long bump may be similar to that in two-stage model. There is not enough information for us to rule out or accept it. Table~\ref{tab:consistency} is a summary of the consistency between the mechanisms for GRB precursors and the observational properties of GRB 221009A. If one property could be interpreted or predicted by the mechanism, the corresponding blank in the table is filled with `Yes'; otherwise, `No' is filled for the inconsistency, and `?' is for the case of no prediction. \textbf{For example, the weak jet is not predicted in the spinar model, thus the blank is filled with `?'. }

\begin{deluxetable*}{l|c|c|c|c|c}
\tabletypesize{\footnotesize}
%\tablewidth{0pt}
\tablecaption{\label{tab:consistency}The consistency between the mechanisms for precursors and GRB 221009A.}
\tablehead{ 
\colhead{GRB 221009A/mechanisms for precursors }
&\colhead{fireball-IS} &\colhead{jet-cocoon interaction} &\colhead{two-stage} &\colhead{magnetar-switch}&\colhead{spinar} 
}
\startdata
Long gap time($\sim100$ s) &No &Yes &Yes &Yes &Yes \\
NT precursor &No &No &Yes &Yes &Yes \\
The jet corresponding to precursor is weak enough &No &No &Yes &No &? \\
The long bump ($\sim60 $ s, $10^{-7}$ erg cm$^{-2}$ s$^{-1}$) before the main burst &No &No &Yes &No &?
%observed  &$\surd$ &$\surd$ &$\surd$ &$\surd$ &? \\
\enddata
\end{deluxetable*}

In general, from the analysis of the first pulse or precursor and the `quiescent' time, it is proposed that the properties of the first pulse are well consistent with the prediction by mechanism for the precursor in the `two-stage' model in the collapsar scenario. Moreover, the first pulse is different from the traditional definition of the precursor because of the weak emission in the gap time. %Here we just discuss its possible origin in a conservative way, which means that the observation could be consistent with the `two-stage' model in the initial collapsar scenario, rather than that an initial collapsar scenario must be inferred from the observation.

\section{discussion and summary}\label{sec:discussion}
 We present an approach to infer the possible ranges of $\Gamma_{\rm b}$ and $L_{\rm j}$ of the jet for the first pulse with the constraints from the duration and flux in a collapsar scenario. Furthermore, this approach could be used to speculate the origins of the precursors of GRBs in the future study.

The first pulse of GRB 221009A is non-thermal, which is the difference between GRB 221009A and GRB 160625B~\citep[e.g.,][]{2018NatAs...2...69Z}; the first precursor of the latter is dominated by a thermal component. In the scenario of GRB 160625B, the precursor occurs after the formation of the accretion disk. As a comparison, we consider that for GRB 221009A the precursor is from the weak jet produced by a rotating PNS during the initial core-collapse phase, rather than the initial prompt accretion phase, as shown in Figure~\ref{fig:centralengine}. Considering the estimated luminosity ($\lesssim10^{49}$ erg s$^{-1}$) with duration of tens of seconds, the total energy ($\sim 10^{50}$ erg) the jet carried is well consistent with that predicted by e.g.~\cite{1970ApJ...161..541L,2000ApJ...537..810W}.  In summary, the origin for the first pulse is discussed conservatively in this analysis, and a weak jet from the initial core-collapse phase in the `two-stage' scenario is taken as the most likely origin, while the other origin for the precursor,  the spinar model, is not ruled out. %The emission mechanism for the main burst may offer a criterion if it is from the jet launched by the BZ mechanism and a 

As the brightest GRB ever detected, GRB 221009A may provide a case that reveals some interesting features which are hidden in those bursts that are not so bright. If the source of the burst had a high $z$, or the observations were not so head-on, the weak emission in the gap time might be missed in the detection. In that case, the gap time seems quiescent and the first pulse should seem similar to the precursors detected before. However, in GRB 221009A, a weak emission during the gap time is observed which enriches the GRB-precursor phenomena, and is important for us to understand the
physical mechanisms of the GRB central engine.

%TC:ignore
\begin{acknowledgements}
 The authors thank supports from the National Program on Key Research and Development Project (2021YFA0718500). This work was partially supported by International Partnership Program of Chinese Academy of Sciences (Grant No.113111KYSB20190020). The authors are very grateful to the GRB data of Fermi/GBM, HXMT and HEBS and konus-Wind.
 We are very grateful for the comments and suggestions of the anonymous referees. 
 Dr. Xin-Ying Song thanks Dr Ming-Yu Ge and Dr Yuan You for the suggestions on the background estimation.
\end{acknowledgements}
\appendix
\section{About background estimation}\label{sec:bg}
\begin{figure*}
\begin{center}
 \centering
   \includegraphics[width=0.8\textwidth]{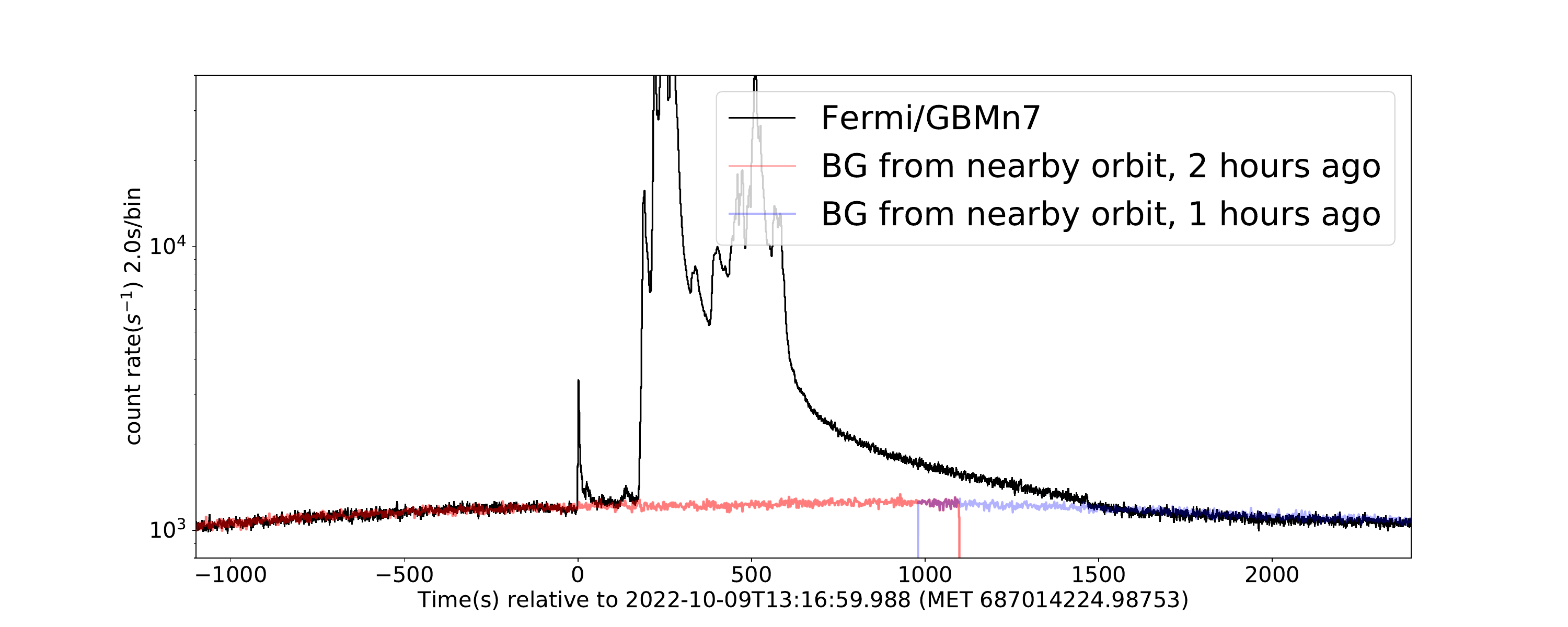}\put(-140,150){(a)}\\
   \includegraphics[width=0.8\textwidth]{Fermi_GBMn8_GBM.pdf}\put(-310,100){(b)}\\
   \includegraphics[width=0.8\textwidth]{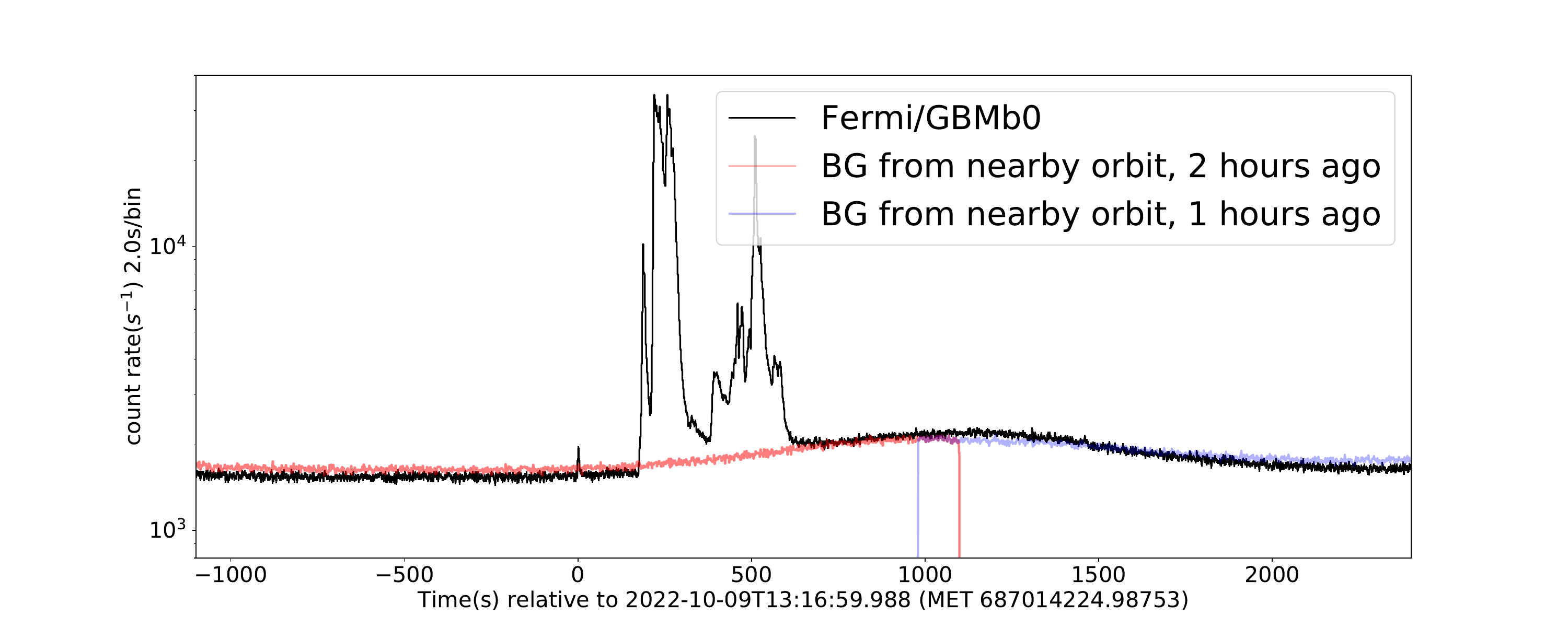}\put(-310,100){(c)}
\caption{The light curves of BGO, NaI 7 and NaI 8 from $T_0-1100$ s to $T_0+2500$ s and nearby orbits.  
\label{fig:bgLC} }
\end{center}
\end{figure*}

\begin{figure*}
\begin{center}
 \centering
   %\includegraphics[width=0.5\textwidth]{PL0.pdf}\put(-140,150){(a)}\\
   \includegraphics[width=0.5\textwidth]{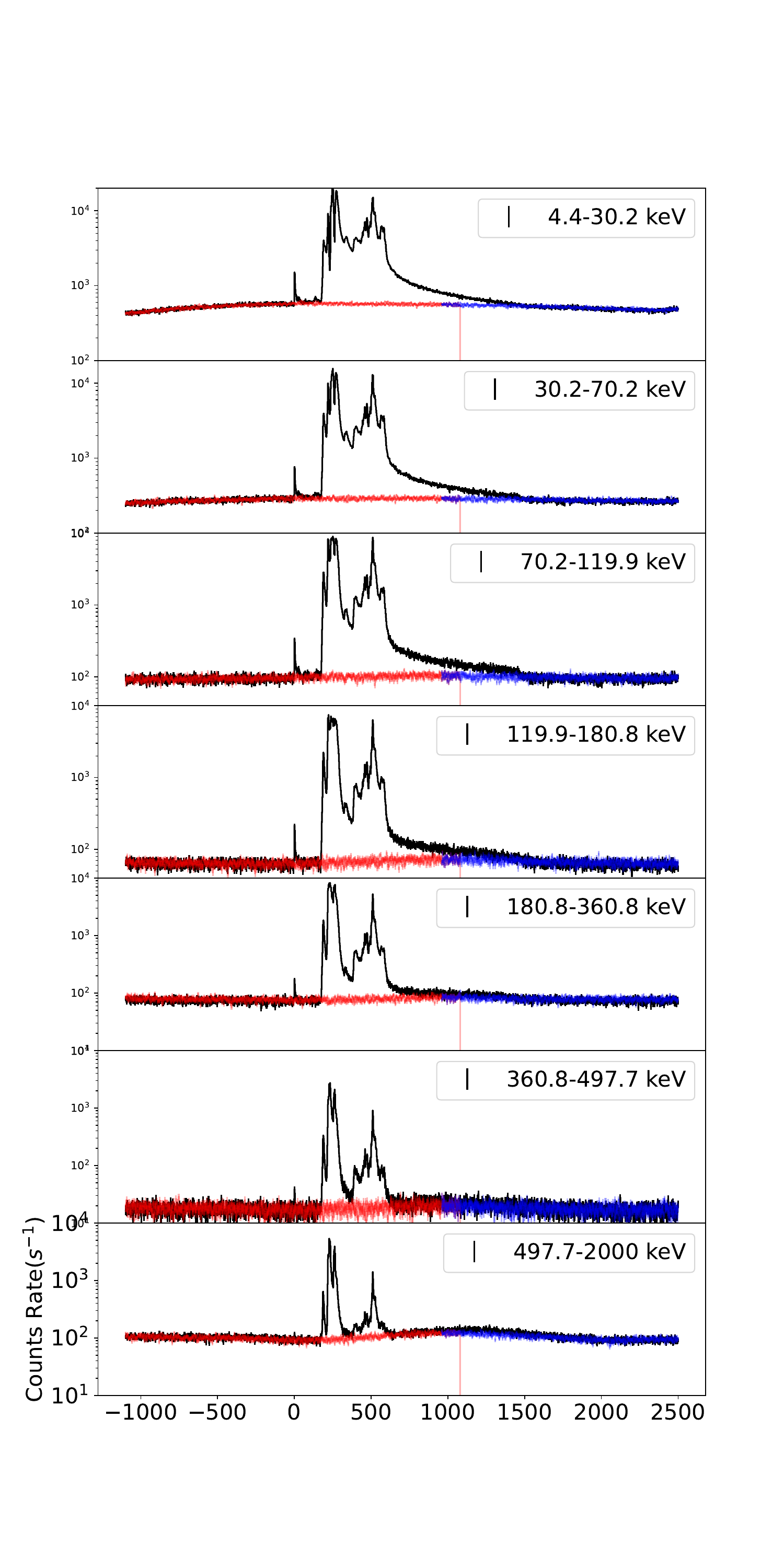}\put(-140,460){(a)  NaI 7 }
   \includegraphics[width=0.5\textwidth]{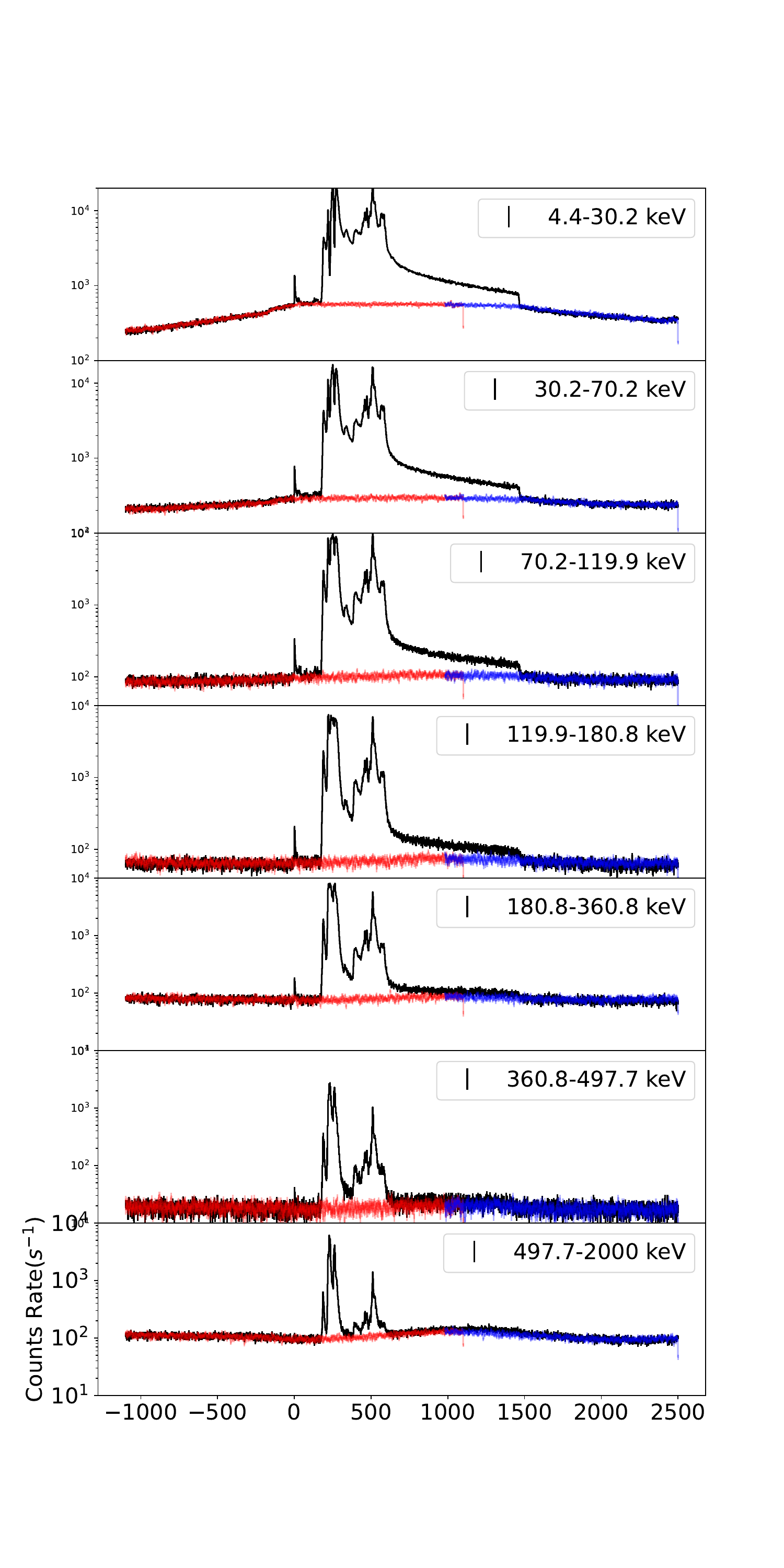}\put(-140,460){(b) NaI 8}
\caption{The light curves of NaI 7, 8 and BGO 0 from $T_0-1100$ s to $T_0+2500$ s and nearby orbits in different energy bands.  
\label{fig:bgLC2} }
\end{center}
\end{figure*}

The background (BG) estimation for extremely long GRB 221009A is important. Figure~\ref{fig:bgLC} shows the events data from the nearby orbits, which helps us know the shape of BG. The shift time ($\sim 5720$ s) is determined by the smallest sum of squares of the difference between the two light curves from $T_0-1000$ s to $T_0$ from GRB 221009A and nearby orbit.  Figure~\ref{fig:bgLC2} shows the light curves and events data from the nearby orbit as BG in different energy bands. The BG could be well described by the data from nearby orbits for NaI 7 and 8 detectors. However, from Figure~\ref{fig:bgLC} (c) and ~\ref{fig:bgLC2} (c) for BGO data, the nearby data are not very consistent with those from trigger of GRB 221009A. From above 385 keV, we find the the GRB events ends at $\sim$600 s. Therefore, we could use a polynomial to describe $T_0$ +[-1100,-5] s and [900, 2400] s, so that the peaking structure could be well described. The contribution below 385.2 keV in BGO 0 (channels 0-5) is ignored in the fitting procedure. 

\begin{figure*}
\begin{center}
 \centering
   %\includegraphics[width=0.5\textwidth]{PL0.pdf}\put(-140,150){(a)}\\
   \includegraphics[width=0.68\textwidth]{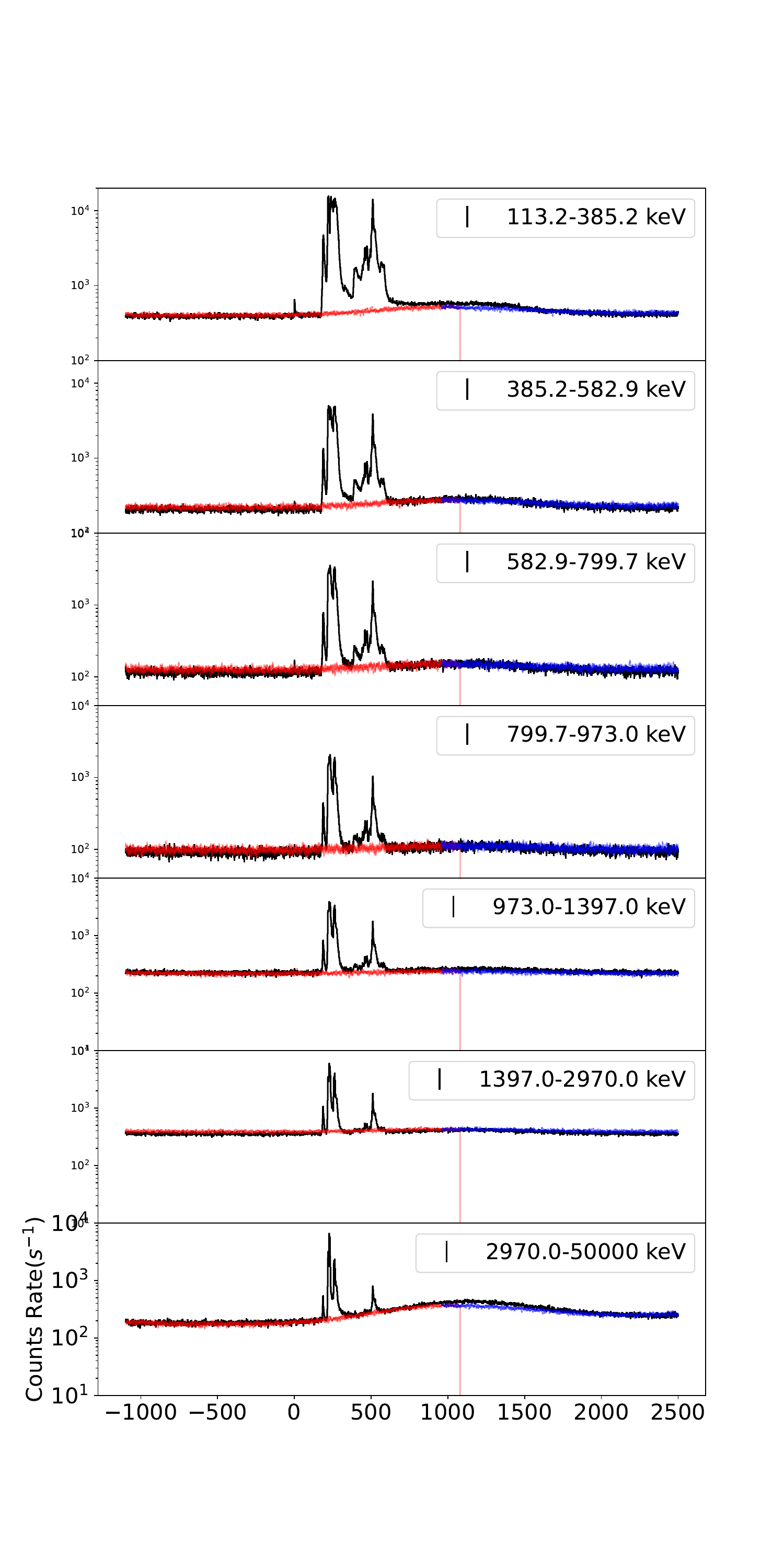}\put(-180,620){(c) BGO 0}
   \center{Figure. \ref{fig:bgLC2}--- Continued.}
%\caption{The light curves of BGO, NaI 7 and 8 from $T_0-1100$ s to $T_0+2500$ s and nearby orbits. }
\end{center}
\end{figure*}
\clearpage

%TC:endignore
\bibliography{GRB221009A}{}

\begin{thebibliography}{}
\expandafter\ifx\csname natexlab\endcsname\relax\def\natexlab#1{#1}\fi
\providecommand{\url}[1]{\href{#1}{#1}}
\providecommand{\dodoi}[1]{doi:~\href{http://doi.org/#1}{\nolinkurl{#1}}}
\providecommand{\doeprint}[1]{\href{http://ascl.net/#1}{\nolinkurl{http://ascl.net/#1}}}
\providecommand{\doarXiv}[1]{\href{https://arxiv.org/abs/#1}{\nolinkurl{https://arxiv.org/abs/#1}}}

\bibitem[{An {et~al.}(2023)An, Antier, Bi, Bu, Cai, Cao, Camisasca, Chang,
  Chen, Chen, Chen, Chen, Chen, Chen, Chen, Coughlin, Cui, Dai,
  Hussenot-Desenonges, Du, Du, Du, Fan, Frontera, Gao, Gao, Ge, Gong, Gu, Guan,
  Guo, Guo, Guidorzi, Han, He, He, Hou, Huang, Huo, Ji, Jia, Jiang, Kann,
  Klotz, Kong, Lan, Li, Li, Li, Li, Li, Li, Li, Li, Li, Li, Li, Li, Li, Liang,
  Liang, Liao, Lin, Liu, Liu, Liu, Liu, Liu, Liu, Liu, Lu, Lu, Lu, Luo, Luo,
  Ma, Ma, Ma, Ma, Maccary, Mao, Meng, Nie, Orlandini, Ou, Peng, Peng, Qiao, Qu,
  Ren, Shi, Shi, Song, Song, Su, Sun, Sun, Sun, Tan, Tan, Tao, Tuo, Turpin,
  Wang, Wang, Wang, Wang, Wang, Wang, Wang, Wang, Wang, Wang, Wang, Wang, Wang,
  Wang, Wen, Wu, Wu, Wu, Xiao, Xiao, Xiao, Xie, Xiong, Xiong, Xu, Xu, Xu, Xu,
  Xu, Xu, Xue, Yang, Yang, Yang, Ye, Yi, Yi, Yin, You, Yu, Yu, Yu, Zeng, Zhang,
  Zhang, Zhang, Zhang, Zhang, Zhang, Zhang, Zhang, Zhang, Zhang, Zhang, Zhang,
  Zhang, Zhang, Zhang, Zhang, Zhang, Zhang, Zhang, Zhao, Zhao, Zhao, Zhao,
  Zhao, Zhao, Zhao, Zhao, Zheng, Zheng, Zhou, Zhou, \& Zhu}]{an2023insighthxmt}
An, Z.-H., Antier, S., Bi, X.-Z., {et~al.} 2023.
\newblock \doarXiv{2303.01203}

\bibitem[{{Bernardini} {et~al.}(2013){Bernardini}, {Campana}, {Ghisellini},
  {D'Avanzo}, {Burlon}, {Covino}, {Ghirlanda}, {Melandri}, {Salvaterra},
  {Vergani}, {D'Elia}, {Fugazza}, {Sbarufatti}, \&
  {Tagliaferri}}]{2013ApJ...775...67B}
{Bernardini}, M.~G., {Campana}, S., {Ghisellini}, G., {et~al.} 2013, \apj, 775,
  67, \dodoi{10.1088/0004-637X/775/1/67}

\bibitem[{{Bi} {et~al.}(2018){Bi}, {Mao}, {Liu}, \&
  {Bai}}]{2018ApJ...866...97B}
{Bi}, X., {Mao}, J., {Liu}, C., \& {Bai}, J.-M. 2018, \apj, 866, 97,
  \dodoi{10.3847/1538-4357/aadcf8}

\bibitem[{{Bissaldi} {et~al.}(2022){Bissaldi}, {Omodei}, {Kerr}, \& {Fermi-LAT
  Team}}]{2022GCN.32637....1B}
{Bissaldi}, E., {Omodei}, N., {Kerr}, M., \& {Fermi-LAT Team}. 2022, GRB
  Coordinates Network, 32637, 1

\bibitem[{{Blandford} \& {Znajek}(1977)}]{1977MNRAS.179..433B}
{Blandford}, R.~D., \& {Znajek}, R.~L. 1977, \mnras, 179, 433,
  \dodoi{10.1093/mnras/179.3.433}

\bibitem[{{Burgess}(2014)}]{2014On}
{Burgess}, J.~M. 2014, \mnras, 445, 2589, \dodoi{10.1093/mnras/stu1925}

\bibitem[{{Burlon} {et~al.}(2009){Burlon}, {Ghirlanda}, {Ghisellini},
  {Greiner}, \& {Celotti}}]{2009A&A...505..569B}
{Burlon}, D., {Ghirlanda}, G., {Ghisellini}, G., {Greiner}, J., \& {Celotti},
  A. 2009, \aap, 505, 569, \dodoi{10.1051/0004-6361/200912662}

\bibitem[{{Cheng} \& {Dai}(2001)}]{2001APh....16...67C}
{Cheng}, K.~S., \& {Dai}, Z.~G. 2001, Astroparticle Physics, 16, 67,
  \dodoi{10.1016/S0927-6505(00)00172-9}

\bibitem[{{de Ugarte Postigo} {et~al.}(2022){de Ugarte Postigo}, {Izzo},
  {Pugliese}, {Xu}, {Schneider}, {Fynbo}, {Tanvir}, {Malesani}, {Saccardi},
  {Kann}, {Wiersema}, {Gompertz}, {Thoene}, {Levan}, \& {Stargate
  Collaboration}}]{2022GCN.32648....1D}
{de Ugarte Postigo}, A., {Izzo}, L., {Pugliese}, G., {et~al.} 2022, GRB
  Coordinates Network, 32648, 1

\bibitem[{{Dichiara} {et~al.}(2022){Dichiara}, {Gropp}, {Kennea}, {Kuin},
  {Lien}, {Marshall}, {Tohuvavohu}, {Williams}, \& {Neil Gehrels Swift
  Observatory Team}}]{2022GCN.32632....1D}
{Dichiara}, S., {Gropp}, J.~D., {Kennea}, J.~A., {et~al.} 2022, GRB Coordinates
  Network, 32632, 1

\bibitem[{{Frederiks} {et~al.}(2023){Frederiks}, {Svinkin}, {Lysenko},
  {Molkov}, {Tsvetkova}, {Ulanov}, {Ridnaia}, {Lutovinov}, {Lapshov},
  {Tkachenko}, \& {Levin}}]{2023arXiv230213383F}
{Frederiks}, D., {Svinkin}, D., {Lysenko}, A.~L., {et~al.} 2023, arXiv
  e-prints, arXiv:2302.13383, \dodoi{10.48550/arXiv.2302.13383}

\bibitem[{{Huang} {et~al.}(2022){Huang}, {Hu}, {Chen}, {Zha}, {Liu}, {Yao},
  {Cao}, \& {Experiment}}]{2022GCN.32677....1H}
{Huang}, Y., {Hu}, S., {Chen}, S., {et~al.} 2022, GRB Coordinates Network,
  32677, 1

\bibitem[{{Klu{\'z}niak} \& {Ruderman}(1998)}]{1998ApJ...505L.113K}
{Klu{\'z}niak}, W., \& {Ruderman}, M. 1998, \apjl, 505, L113,
  \dodoi{10.1086/311622}

\bibitem[{{Lazzati}(2005)}]{2005MNRAS.357..722L}
{Lazzati}, D. 2005, \mnras, 357, 722, \dodoi{10.1111/j.1365-2966.2005.08687.x}

\bibitem[{{LeBlanc} \& {Wilson}(1970)}]{1970ApJ...161..541L}
{LeBlanc}, J.~M., \& {Wilson}, J.~R. 1970, \apj, 161, 541,
  \dodoi{10.1086/150558}

\bibitem[{{Lesage} \& {Fermi Gamma-ray Burst Monitor
  Team}(2022)}]{2022GCN.31565....1L}
{Lesage}, S., \& {Fermi Gamma-ray Burst Monitor Team}. 2022, GRB Coordinates
  Network, 31565, 1

\bibitem[{{Li} \& {Mao}(2022)}]{2022ApJ...928..152L}
{Li}, L., \& {Mao}, J. 2022, \apj, 928, 152, \dodoi{10.3847/1538-4357/ac4af2}

\bibitem[{{Li} {et~al.}(2019){Li}, {Geng}, {Meng}, {Wu}, {Huang}, {Wang},
  {Moradi}, {Uhm}, \& {Zhang}}]{2019ApJ...884..109L}
{Li}, L., {Geng}, J.-J., {Meng}, Y.-Z., {et~al.} 2019, \apj, 884, 109,
  \dodoi{10.3847/1538-4357/ab40b9}

\bibitem[{{Lipunov} \& {Gorbovskoy}(2007)}]{2007ApJ...665L..97L}
{Lipunov}, V., \& {Gorbovskoy}, E. 2007, \apjl, 665, L97,
  \dodoi{10.1086/521099}

\bibitem[{{Lipunova} {et~al.}(2009){Lipunova}, {Gorbovskoy}, {Bogomazov}, \&
  {Lipunov}}]{2009MNRAS.397.1695L}
{Lipunova}, G.~V., {Gorbovskoy}, E.~S., {Bogomazov}, A.~I., \& {Lipunov}, V.~M.
  2009, \mnras, 397, 1695, \dodoi{10.1111/j.1365-2966.2009.15079.x}

\bibitem[{{Liu} {et~al.}(2022){Liu}, {Zhang}, {Xiong}, {Zheng}, {Wang}, {Xue},
  {Qiao}, {Tan}, {Zhang}, {Li}, {Wen}, {Peng}, {Song}, {Zheng}, {Guo}, {Li},
  {Ma}, {Huang}, {Zhao}, {Wang}, {Wang}, {Zhang}, {Du}, {Liang}, {Lu}, {Wu},
  {Yu}, {Xiao}, {Cai}, {Zhang}, {Li}, {An}, {Gao}, {Gong}, {Liu}, {Liu}, {Sun},
  {Xu}, {Yang}, {Feng}, {Wang}, {Zhang}, {Chen}, {Lu}, {Zhang}, {Gecam}, \&
  {Hebs Teams}}]{2022GCN.32751....1L}
{Liu}, J.~C., {Zhang}, Y.~Q., {Xiong}, S.~L., {et~al.} 2022, GRB Coordinates
  Network, 32751, 1

\bibitem[{{MacFadyen} {et~al.}(2001){MacFadyen}, {Woosley}, \&
  {Heger}}]{2001ApJ...550..410M}
{MacFadyen}, A.~I., {Woosley}, S.~E., \& {Heger}, A. 2001, \apj, 550, 410,
  \dodoi{10.1086/319698}

\bibitem[{{M{\'e}sz{\'a}ros} \& {Rees}(2000)}]{2000ApJ...530..292M}
{M{\'e}sz{\'a}ros}, P., \& {Rees}, M.~J. 2000, \apj, 530, 292,
  \dodoi{10.1086/308371}

\bibitem[{{Nakar} \& {Piran}(2017)}]{2017ApJ...834...28N}
{Nakar}, E., \& {Piran}, T. 2017, \apj, 834, 28,
  \dodoi{10.3847/1538-4357/834/1/28}

\bibitem[{{Preece} {et~al.}(1998){Preece}, {Briggs}, {Mallozzi}, {Pendleton},
  {Paciesas}, \& {Band}}]{Preece_1998}
{Preece}, R.~D., {Briggs}, M.~S., {Mallozzi}, R.~S., {et~al.} 1998, \apjl, 506,
  L23, \dodoi{10.1086/311644}

\bibitem[{{Ramirez-Ruiz} {et~al.}(2002){Ramirez-Ruiz}, {MacFadyen}, \&
  {Lazzati}}]{2002MNRAS.331..197R}
{Ramirez-Ruiz}, E., {MacFadyen}, A.~I., \& {Lazzati}, D. 2002, \mnras, 331,
  197, \dodoi{10.1046/j.1365-8711.2002.05176.x}

\bibitem[{{Ren} {et~al.}(2022){Ren}, {Wang}, \& {Zhang}}]{2022arXiv221010673R}
{Ren}, J., {Wang}, Y., \& {Zhang}, L.-L. 2022, arXiv e-prints,
  arXiv:2210.10673, \dodoi{10.48550/arXiv.2210.10673}

\bibitem[{{Scargle} {et~al.}(2013){Scargle}, {Norris}, {Jackson}, \&
  {Chiang}}]{2013ApJ...764..167S}
{Scargle}, J.~D., {Norris}, J.~P., {Jackson}, B., \& {Chiang}, J. 2013, \apj,
  764, 167, \dodoi{10.1088/0004-637X/764/2/167}

\bibitem[{{Siegel} {et~al.}(2014){Siegel}, {Ciolfi}, \&
  {Rezzolla}}]{2014ApJ...785L...6S}
{Siegel}, D.~M., {Ciolfi}, R., \& {Rezzolla}, L. 2014, \apjl, 785, L6,
  \dodoi{10.1088/2041-8205/785/1/L6}

\bibitem[{{Song} {et~al.}(2022{\natexlab{a}}){Song}, {Zhang}, {Ge}, \&
  {Zhang}}]{2022MNRAS.517.2088S}
{Song}, X.-Y., {Zhang}, S.-N., {Ge}, M.-Y., \& {Zhang}, S. 2022{\natexlab{a}},
  \mnras, 517, 2088, \dodoi{10.1093/mnras/stac2764}

\bibitem[{{Song} {et~al.}(2022{\natexlab{b}}){Song}, {Xiong}, {Zhang}, {Li},
  {Li}, {Huang}, {Guidorzi}, {Frontera}, {Liu}, {Li}, {Li}, {Liao}, {Cai},
  {Luo}, {Xiao}, {Yi}, {Zheng}, {Zhou}, {Liu}, {Xue}, {Zhang}, {Zheng},
  {Chang}, {Li}, {Lu}, {Zhang}, {Zhang}, {Jin}, {Li}, {Lu}, {Song}, {Wu}, {Xu},
  {Ma}, {Ge}, {Jia}, {Li}, {Nie}, {Wang}, {Zhang}, {Zheng}, {Yang}, \&
  {Yang}}]{2022ApJS..259...46S}
{Song}, X.-Y., {Xiong}, S.-L., {Zhang}, S.-N., {et~al.} 2022{\natexlab{b}},
  \apjs, 259, 46, \dodoi{10.3847/1538-4365/ac4d22}

\bibitem[{{Svinkin} {et~al.}(2022){Svinkin}, {Frederiks}, {Ulanov},
  {Tsvetkova}, {Lysenko}, {Ridnaia}, {Cline}, \& {Konus-Wind
  Team}}]{2022GCN.31604....1S}
{Svinkin}, D., {Frederiks}, D., {Ulanov}, M., {et~al.} 2022, GRB Coordinates
  Network, 31604, 1

\bibitem[{{Tan} {et~al.}(2022){Tan}, {Li}, {Ge}, {Li}, {Xiong}, \&
  {Zhang}}]{2022ATel15660....1T}
{Tan}, W.~J., {Li}, C.~K., {Ge}, M.~Y., {et~al.} 2022, The Astronomer's
  Telegram, 15660, 1

\bibitem[{{Uhm} \& {Zhang}(2014)}]{2014NatPh..10..351U}
{Uhm}, Z.~L., \& {Zhang}, B. 2014, Nature Physics, 10, 351,
  \dodoi{10.1038/nphys2932}

\bibitem[{{Wang} \& {M{\'e}sz{\'a}ros}(2007)}]{2007ApJ...670.1247W}
{Wang}, X.-Y., \& {M{\'e}sz{\'a}ros}, P. 2007, \apj, 670, 1247,
  \dodoi{10.1086/522820}

\bibitem[{{Wei} {et~al.}(2016){Wei}, {Wu}, \& {Melia}}]{2016MNRAS.463.1144W}
{Wei}, J.-J., {Wu}, X.-F., \& {Melia}, F. 2016, \mnras, 463, 1144,
  \dodoi{10.1093/mnras/stw2057}

\bibitem[{{Wheeler} {et~al.}(2000){Wheeler}, {Yi}, {H{\"o}flich}, \&
  {Wang}}]{2000ApJ...537..810W}
{Wheeler}, J.~C., {Yi}, I., {H{\"o}flich}, P., \& {Wang}, L. 2000, \apj, 537,
  810, \dodoi{10.1086/309055}

\bibitem[{{Zhang} \& {Yan}(2011)}]{2011ApJ...726...90Z}
{Zhang}, B., \& {Yan}, H. 2011, \apj, 726, 90,
  \dodoi{10.1088/0004-637X/726/2/90}

\bibitem[{{Zhang} {et~al.}(2018){Zhang}, {Zhang}, {Castro-Tirado}, {Dai},
  {Tam}, {Wang}, {Hu}, {Karpov}, {Pozanenko}, {Zhang}, {Mazaeva}, {Minaev},
  {Volnova}, {Oates}, {Gao}, {Wu}, {Shao}, {Tang}, {Beskin}, {Biryukov},
  {Bondar}, {Ivanov}, {Katkova}, {Orekhova}, {Perkov}, {Sasyuk}, {Mankiewicz},
  {{\.Z}arnecki}, {Cwiek}, {Opiela}, {Zadro{\.Z}ny}, {Aptekar}, {Frederiks},
  {Svinkin}, {Kusakin}, {Inasaridze}, {Burhonov}, {Rumyantsev}, {Klunko},
  {Moskvitin}, {Fatkhullin}, {Sokolov}, {Valeev}, {Jeong}, {Park},
  {Caballero-Garc{\'\i}a}, {Cunniffe}, {Tello}, {Ferrero}, {Pandey},
  {Jel{\'\i}nek}, {Peng}, {S{\'a}nchez-Ram{\'\i}rez}, \&
  {Castell{\'o}n}}]{2018NatAs...2...69Z}
{Zhang}, B.~B., {Zhang}, B., {Castro-Tirado}, A.~J., {et~al.} 2018, Nature
  Astronomy, 2, 69, \dodoi{10.1038/s41550-017-0309-8}

\end{thebibliography}
\bibliographystyle{aasjournal}
\end{document}